\documentclass[conference]{IEEEtran}

\usepackage{graphicx}
\usepackage{amsmath}
\usepackage{amssymb}
\usepackage{algorithm}
\usepackage{algorithmicx}
\usepackage{algpseudocode}
\usepackage{amsfonts,amssymb,amsthm}
\usepackage{xspace}
\usepackage[acronym,nomain]{glossaries}
\usepackage{caption}
\usepackage{listings}
\usepackage{blindtext}
\usepackage{svn-multi}
\usepackage{multirow}
\usepackage{lipsum}
\usepackage{tikz}
\usetikzlibrary{patterns}
\usepackage{subcaption}
\usepackage{hyperref}
\newcount\outlineon
\newcommand{\outline}[1]{\ifnum\outlineon>0\
	\\\noindent\fbox{\begin{minipage}{\linewidth}{\footnotesize#1}\end{minipage}}\\\\\fi}
\newcommand{\execlude}[1]{}

\newcommand{\proofend}[2]{ $ \blacksquare$ \vspace{5mm} }

\theoremstyle{plain}

\theoremstyle{definition}

\theoremstyle{remark}

\newif\ifcomments

\commentstrue 
\commentsfalse 

\hypersetup{
	bookmarks=true,         
	unicode=false,          
	pdftoolbar=true,        
	pdfmenubar=true,        
	pdffitwindow=false,     
	pdftitle={paper},    
	pdfauthor={Dr. Gautam Gala},     
	pdfsubject={OSPERT paper},   
	pdfcreator={Dr. Gautam Gala},   
	pdfproducer={},  
	pdfnewwindow=true,      
	colorlinks=false,       
	linkcolor=red,          
	citecolor=green,        
	filecolor=magenta,      
	urlcolor=cyan           
}

\ifcomments
	\usepackage[colorinlistoftodos]{todonotes}
\else
	\usepackage[disable]{todonotes}
\fi

\newcommand{\ggcomment}[1]{\todo[inline, color=green!40, author=GG]{#1}}

\makeglossaries
\IEEEoverridecommandlockouts
\begin{document}
	\newacronym{mcs}{MCS}{Mixed-Criticality Systems}
\newacronym{asils}{ASILs}{Automotive Safety and Integrity Levels}
\newacronym{dals}{DALs}{Design/Development Assurance Levels}
\newacronym{sils}{SILs}{Safety Integrity Levels}
\newacronym{ife}{IFE}{In-Flight Entertainment}
\newacronym{fms}{FMS}{Flight Management System}
\newacronym{ima}{IMA}{Integrated modular avionics}
\newacronym{qos}{QoS}{Quality of Service}
\newacronym{dlt}{DLT}{Distributed Ledger Technology}
\newacronym{pbft}{PBFT}{Practical Byzantine Fault-tolerant}
\newacronym[plural=LRMs,firstplural=Local Resource Managers (LRMs)]{lrm}{LRM}{Local Resource Manager}
\newacronym{grm}{GRM}{Global Resource Manager}
\newacronym[plural=DGRMs,firstplural=Distributed Global Resource Managers (DGRMs)]{dgrm}{DGRM}{Distributed Global Resource Manager}
\newacronym{cep}{CEP}{Complex Event Proccessing}
\newacronym[plural=RMEs,firstplural=Resource Management Entities (RMEs)]{rme}{RME}{Resource Management Entity}
\newacronym[plural=RMs,firstplural=Resource Managers (RMs)]{rm}{RM}{Resource Manager}
\newacronym[plural=MONs,firstplural=Local Resource Monitors (MONs)]{mon}{MON}{Local Resource Monitor}
\newacronym[plural=LRSs,firstplural=Local Resource Schedulers (LRSs)]{lrs}{LRS}{Local Resource Scheduler}
\newacronym[plural=TPs,firstplural=Transaction Processors (TPs)]{tp}{TP}{Transaction Processor}
\newacronym[plural=VMs,firstplural=Virtual Machines (VMs)]{vm}{VM}{Virtual Machine}
\newacronym{swap}{SWaP}{Size, Weight and Power}
\newacronym[plural=CLIs,firstplural=Client (CLIs)]{cli}{CLI}{Client}
\newacronym{SC}{SC}{Safety-Critical}
\newacronym{BE}{BE}{Best Effort}
\newacronym{KVM}{KVM}{Kernel Virtual Machine}
\newacronym{TT}{TT}{Time-Triggered}
\newacronym{ET}{ET}{Event-Triggered}
\newacronym{QoS}{QoS}{Quality-of-Service}
\newglossaryentry{VM}{
	type=\acronymtype,
	name={VM},
	first={Virtual Machine},
	firstplural={Virtual Machines (VMs)},
	text={VM},
	plural={VMs},
	description={Virtual Machine}
}
	\ifcomments
	\outlineon = 1
	\fi
	
	%
	\title{Joint Time-and Event-Triggered Scheduling\\in the Linux Kernel 
	}
		
	\author{
		\IEEEauthorblockN{Gautam Gala, Isser Kadusale, and Gerhard Fohler}
		\IEEEauthorblockA{
			University of Kaiserslautern-Landau (RPTU), 
			Germany\\
			\{gala, kadusale, fohler\}@eit.uni-kl.de}
	}
		
	\maketitle

\ggcomment{Overall 8 pages}
\svnid{$Id$}
\begin{abstract}
There is increasing interest in using Linux in the real-time domain due to the emergence of cloud and edge computing, the need to decrease costs, and the growing number of complex functional and non-functional requirements of real-time applications. Linux presents a valuable opportunity as it has rich hardware support, an open-source development model, a well-established programming environment, and avoids vendor lock-in. Although Linux was initially developed as a general-purpose operating system, some real-time capabilities have been added to the kernel over many years to increase its predictability and reduce its scheduling latency. Unfortunately, Linux currently has no support for time-triggered (TT) scheduling, which is widely used in the safety-critical domain for its determinism, low run-time scheduling latency, and strong isolation properties. We present an enhancement of the Linux scheduler as a new low-overhead TT scheduling class to support offline table-driven scheduling of tasks on multicore Linux nodes. Inspired by the Slot shifting algorithm, we complement the new scheduling class with a low overhead slot shifting manager running on a non-time-triggered core to provide guaranteed execution time to real-time aperiodic tasks by using the slack of the time-triggered tasks and avoiding high-overhead table regeneration for adding new periodic tasks. Furthermore, we evaluate our implementation on server-grade hardware with Intel Xeon Scalable Processor.
\end{abstract}


\section{Introduction}
\svnid{$Id$}
\label{section:introduction}

Real-time (RT) Safety-critical industries such as Thales (railway, \cite{gala-railway}) and Airbus (aerospace, \cite{ankit_avionics}) are exploring commercial-off-the-shelf (COTS) hardware platforms to benefit from reduced time to market, lower SWaP and higher computational power. Linux is interesting for COTS platforms due to extensive industrial and academic research, rich COTS hardware support, open-source versatility, and flexibility. It is free to run, study, modify to fit special requirements, and port to different hardware platforms. 
There are many useful open-source or freely available libraries, development tools, and applications created for Linux. Using Linux can help to keep costs down. 
Developers and system designers from all domains find the ability to adapt Linux systems to their requirements and hardware platform advantageous.
Linux has a large open-source community that freely publishes bugs (possibly with fixes) and enhancements to Linux for all to benefit and a possibility for future maintenance by the community. Thus, using Linux often ensures we save efforts to re-invent what already exists.

Linux (in conjunction with KVM) is used to power most of the public cloud infrastructure and is heavily used in supercomputing \cite{Canonical}. With the advent of Industry 4.0, smart healthcare devices, software-defined vehicles, and other cloud-connected transportation (e.g., trains and trams), OEMs of safety-critical domains will find using a single Linux operating systems for cloud/edge servers, IoT/smart devices, and in-vehicle software beneficial. A single operating system will help reduce the costs and the burden of managing multiple suppliers, maintaining many development environments, and numerous application versions. 

Using existing, often proprietary, RT operating systems (RTOSs) or hypervisors is expensive, lacks flexibility and rich hardware support of Linux, can lead to vendor lock-in, and needs tedious re-implementing of existing libraries, tools, and applications for these RTOSs.

Safety-critical industries have considered using RT-capable Linux, e.g., NASA for ground and space operations \cite{nasa}, Thales for an RT-capable cloud to support railway operations \cite{gala-railway}, and various automotive companies for Linux-based fully open software stack for the connected car \cite{sivakumar2022automotive}.  
The Linux community has made significant efforts to support RT applications, e.g., PREEMPT\_RT \cite{preemptrt} and Xenomai \cite{gerum2004xenomai}. These efforts have focused on improving determinism and reducing latency in the Linux Kernel \cite{reghenzani2019real}. The current Linux kernel scheduler supports Constant Bandwidth Server (CBS) based resource reservations over an Earliest-Deadline First (EDF) scheduler to improve RT guarantees \cite{lelli2016deadline}.

Unfortunately, the Linux kernel scheduler still does not support (table-driven) time-triggered (TT) scheduling. Safety-critical RT systems often require TT scheduling. 
Since the creation and validation of the scheduling table occur offline in the TT schedule, a system designer can factor in complex constraints such as latency and precedence constraints, which would otherwise incur large overheads to handle directly at run time. TT scheduling also enforces strong temporal isolation between tasks. It can easily ensure that a task overrunning its worst-case execution time (WCET) does not hamper other tasks' schedulability. A system using TT scheduling is highly predictable as events occur pre-planned at fixed points in time. Thus, testing and certifying the system is easier as there are only a few predictable scenarios to consider.

A TT scheduler mainly consists of a dispatcher that assigns resource(s) to task(s) based on an offline computed schedule (during the design phase). The scheduler receives this schedule as a \textit{scheduling table} consisting of all the necessary scheduling decisions and the point in time the dispatcher should implement those decisions. Since a TT scheduler is quite simple, it has much lower overheads as compared to event-triggered (ET) schedulers (e.g., EDF-based), especially during peak load conditions \cite{kopetz1991event}. Moreover, ET schedulers can produce widely different schedules for the same system when the sequence or timing of events changes, leading to lower predictability. A system with an ET scheduler requires exhaustive testing using simulated loads, considering even the rarest events. However, proving that the tests covered all possible scenarios that may occur at runtime is not easy.

In this paper, we build upon some basic ideas described in \cite{9622384} and propose a new TT CPU scheduler for Linux. The main contributions are:
\begin{enumerate}	
	\item For ensuring static scheduling guarantees and minimizing scheduling latencies, we provide the ability to schedule periodic tasks (Linux threads, processes, VMs, or containers) in Linux using TT (table-driven) scheduling.
	\item Inspired by Slot-shifting \cite{slotshift}, we combine the flexibility to execute RT aperiodic (AP) and best effort (BE) tasks at run time in the slack of the periodic TT tasks, and thus, ensure efficient utilization of resources. We provide the option to add new periodic tasks to the offline table without the need for a high-overhead scheduling table regeneration. 
\end{enumerate}

We implement our approach as a new low-overhead Linux scheduling class (SCHED\_TT) with a separate Slot-Shifting Manager (SSM) Kernel Module to help integrate RT AP tasks. We present an overview of the actual implementation and provide the scheduling overheads by performing experiments on a COTS server with Intel Xeon Processor (Cascade Lake).

The proposed TT CPU scheduler for Linux can be easily integrated with existing real-time resource management and orchestration frameworks to give static scheduling guarantees to tasks, VMs \cite{gala-railway,DGRM}, and containers \cite{monaco2023extensions}.

\section{Related Work}
\subsection{Real-time and Linux}
PREEMPT\_RT~\cite{preemptrt}, an open-source patch for Linux Kernel, is generally used to create real-time Linux. The patch focuses on providing mechanisms to reduce latency and increase determinism in the Linux kernel and complements existing Linux scheduling. Many of the patches are now part of the mainstream Linux kernel. We use a Linux kernel with PREEMPT\_RT as a base for further development.

Efforts from ACTORS EU project~\cite{faggioli2009edf} followed by the work from Lelli et al.~\cite{lelli2016deadline}, and others led to the introduction of SCHED\_DEADLINE in the Linux kernel scheduler to support CBS-based resource reservations over an EDF scheduler.

Litmus$^{RT}$ provides a real-time extension for the Linux kernel to act as an experimental testbed for real-time research, especially on multiprocessor real-time scheduling and synchronization. It supports various clustered, partitioned, global, and semi-partitioned schedulers such as partitioned-EDF and partitioned fixed-priority scheduling. 

Approaches such as RTAI~\cite{arm2016real}, Xenomai~\cite{gerum2004xenomai}, and RTLinux~\cite{yodaiken1999rtlinux} use a hardware abstraction layer below the Linux kernel. The Linux kernel runs as the lowest priority (background) thread on top of this layer. Similarly, the Jailhouse partitioning hypervisor~\cite{baryshnikov2016jailhouse} runs bare metal and cooperates closely with Linux to run safety-critical applications. However, guests cannot share a CPU because Jailhouse has no scheduler.
None of these approaches support running TT tasks natively by Linux.

\subsection{Existing TT scheduling support in OSes or hypervisors}
TTTech MotionWise Platform~\cite{motionwise} emulates a TT scheduler in userspace via standard POSIX system calls. However, this solution has high scheduling latencies and jitter compared to a kernel-level implementation~\cite{schedTTech}. 

Specialized real-time hypervisors, such as XtratuM~\cite{xtratum} and PikeOS~\cite{pikeos}, support TT schedule. However, they are often proprietary and have limited hardware and software support. Moreover, they are unsuitable for cloud environments as they support only a handful of guest operating systems and have additional limitations, such as limited CPU models for VMs.

Xen with the ARINC 653 scheduler~\cite{xenarinc} only considers cyclic scheduling on a single core.
The Tableau~\cite{tableau} extension to the Xen hypervisor introduces support for TT scheduling of VMs to reduce scheduling overheads and enable high-density packaging of VMs. However, Tableau is not explicitly targeted to execute periodic safety-critical real-time VMs and uses a table regeneration process to add new tasks. 
Gala et al.~\cite{gala-railway} demonstrated that Xen has, in general, higher overheads than Linux (+ PREEMPT\_RT patch) in conjunction with KVM. Therefore, KVM/Linux is better suited for low latency RT applications than Xen.

Very recently, Karachatzis et al.~\cite{schedTTech} presented an implementation and evaluation of a kernel-level time-triggered scheduling approach for Linux. 
The approach proposed improving the node's utilization by allowing other tasks to run in the slack of TT tasks. In addition to allowing non-RT tasks in the slack of TT tasks, our approach supports the guaranteed execution of new RT tasks by appending them to the scheduling table at run-time (without needing to regenerate the table).  Our approach takes advantage of the slot-shifting algorithm to add the required flexibility. Our approach keeps run-time overhead low by separating the TT task dispatching via newly created SCHED\_TT scheduling class (on TT cores) from the slot-shifting algorithm decision-making. The slot-shifting algorithm is executed on non-TT cores with a period equal to the slot length.


\subsection{Joint scheduling of TT and ET RT tasks}
Many previous scheduling algorithms have explored combining both time-and event-triggered approaches to take advantage of both the contrary scheduling approaches.
Fohler~\cite{slotshift} presented the Slot shifting algorithm for joint scheduling of TT and ET tasks. Schorr~\cite{schorr2010online} presented a multiprocessor extension to the slot-shifting algorithm. However, previous works do not integrate these and other approaches (e.g.,~\cite{real,syed2018job}) to combine TT and ET tasks with the Linux kernel scheduler.

\section{Background}
\subsection{Slot Shifting}
Let us consider a global time whose progress is triggered by equidistant events. We consider a sparse time base where the time duration separating two of these events is called a \textit{slot}
A multicore slot on a cloud node can run $N$ tasks at a time, where $N$ is the total number of CPU cores scheduled under the TT schedule. For simplicity of explanation, we assume each task requires only one core. Thus, a slot consists of CPU allocation in time units (e.g., milliseconds) and a mapping of tasks to TT cores.
We used an existing heuristic algorithm to create the mapping in the form of an offline scheduling table.
The heuristic must know all tasks in the system before run time to create the scheduling table. 
We must store the scheduling table in the node. The node's scheduler uses it to execute tasks once the node boots.
At run time, the TT scheduler cannot handle a task that is not present in the offline scheduling table. 

Since the exact activation times of sporadic or aperiodic RT tasks are unknown before run time, we must consider them periodic when creating the scheduling table. Moreover, if the system designers need to add a new task to an existing scheduling table, they must recompute it. While creating the scheduling table, the task resource assignment occurs based on the worst-case resource demand. In the average case, most tasks utilize only a fraction of the allocated resources at run time.

The slot-Shifting \cite{slotshift, schorr2010online} algorithm provides a way to run or add AP RT tasks in the slack of TT tasks at run time without needing an expensive table regeneration process.
Slot shifting defines \textit{capacity interval} (or simply \textit{interval}) as a set of consecutive slots that possess the exact mapping of tasks to cores. In other words, a new interval starts on a slot when the set of tasks assigned to this slot differs from the previous one.
Intervals may contain some cores without any task assignment, resulting in idle slots within the interval. These slots form the \textit{Spare Capacity (SC)} of the interval. The scheduler uses SC to execute AP RT tasks with guaranteed CPU execution time. 

The scheduler executes any task assigned to the current interval (as per the scheduling table). In idle slots, it runs any ready tasks defined in the scheduling table that do not belong to the current interval. If no task is ready, the scheduler executes a best-effort or idle task. The scheduler reduces the run time SC at the end of every slot where a task assigned to the current interval does not execute. The run time SC increases when a task assigned to the current interval finishes before its WCET.

Slot shifting define \textit{acceptance test} and a \textit{guarantee routine} to handle AP RT tasks \cite{schorr2010online}.
When a new AP RT task arrives, the acceptance test determines the sum of the remaining SC in the current interval, spare capacities in all the intervals before the task's deadline, and the usable SC in the interval where the deadline of the AP RT task lies.
Then it checks if the determined sum is more significant than the WCET of the task (in terms of slots).
If the test is successful, the guarantee routine adds the task temporarily to the scheduling table and updates the spare capacities of all affected intervals. The guarantee routine may need to split existing intervals. The result is that this routine guarantees the CPU allocation to the newly accepted AP RT task. Using a similar idea, we can permanently add a new periodic RT task to the scheduling table without requiring a high-overhead table regeneration process by considering offline SC values (instead of online values).

\subsection{Linux scheduler}
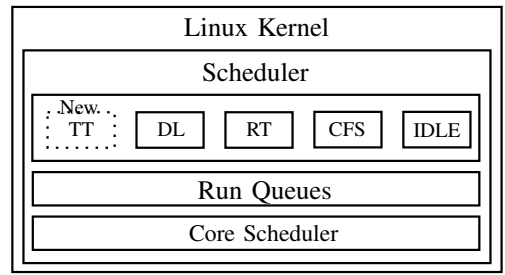
\begin{figure}[b]
	\centering
	\resizebox{0.8\linewidth}{!} {
		\tikzset{every picture/.style={line width=0.75pt}} 

\begin{tikzpicture}[x=0.75pt,y=0.75pt,yscale=-1,xscale=1]

\draw   (65.17,0.65) -- (285.17,0.65) -- (285.17,119.95) -- (65.17,119.95) -- cycle ;
\draw   (70.17,20.2) -- (280.17,20.2) -- (280.17,115.95) -- (70.17,115.95) -- cycle ;
\draw   (74.17,40.2) -- (275.17,40.2) -- (275.17,69.95) -- (74.17,69.95) -- cycle ;
\draw   (74.37,95) -- (275.17,95) -- (275.17,110) -- (74.37,110) -- cycle ;
\draw   (74.37,75) -- (275.17,75) -- (275.17,90) -- (74.37,90) -- cycle ;
\draw   (121,48.05) -- (151.17,48.05) -- (151.17,64.03) -- (121,64.03) -- cycle ;
\draw   (161,48.05) -- (191.17,48.05) -- (191.17,64.03) -- (161,64.03) -- cycle ;
\draw   (201,47.88) -- (231.17,47.88) -- (231.17,63.87) -- (201,63.87) -- cycle ;
\draw   (241,47.88) -- (271.17,47.88) -- (271.17,63.87) -- (241,63.87) -- cycle ;
\draw  [dash pattern={on 0.84pt off 2.51pt}] (81,48.05) -- (111.17,48.05) -- (111.17,64.03) -- (81,64.03) -- cycle ;

\draw (141,4.2) node [anchor=north west][inner sep=0.75pt]  [font=\small] [align=left] {Linux Kernel};
\draw (149,24.2) node [anchor=north west][inner sep=0.75pt]  [font=\small] [align=left] {Scheduler};
\draw (129,51.2) node [anchor=north west][inner sep=0.75pt]   [align=left] {{\scriptsize DL}};
\draw (244,51.45) node [anchor=north west][inner sep=0.75pt]   [align=left] {{\scriptsize IDLE}};
\draw (206,51.03) node [anchor=north west][inner sep=0.75pt]   [align=left] {{\scriptsize CFS}};
\draw (169,51.2) node [anchor=north west][inner sep=0.75pt]   [align=left] {{\scriptsize RT}};
\draw (147.12,77.2) node [anchor=north west][inner sep=0.75pt]  [font=\footnotesize] [align=left] {{\small Run Queues}};
\draw (143.12,97) node [anchor=north west][inner sep=0.75pt]  [font=\small] [align=left] {{\footnotesize Core Scheduler }};
\draw (89,51.2) node [anchor=north west][inner sep=0.75pt]   [align=left] {{\scriptsize TT}};
\draw (85,41.2) node [anchor=north west][inner sep=0.75pt]   [align=left] {{\scriptsize New}};

\end{tikzpicture}
	}
	\caption{Relevant parts of Linux CPU Scheduler}	
	\label{fig:scheduler}
\end{figure}

The Linux scheduler is part of the kernel that helps manage tasks and decide which to run next. It is modular and designed to be extensible \cite{LinuxModularScheduler}. It is a multi-queue scheduler that maintains a per-core run queue of ready tasks. It consists of a core scheduler and extensible modules called the scheduling classes, as shown in figure \ref{fig:scheduler}. Each class encapsulates a scheduling policy to decide which ready task (belonging to that class) to run next (from the run queue).  

Linux currently supports four main scheduling classes: 
\begin{enumerate}
	\item \textit{Deadline (DL)} for SCHED\_DEADLINE policy to support CBS-based resource reservations over EDF scheduling.
	\item \textit{Real-Time (RT)} for POSIX fixed-priority scheduling with SCHED\_RR (round-robin) and SCHED\_FIFO policies.
	\item \textit{Completely Fair Scheduling (CFS)} to maintain balance in allocating processor time to tasks.
	\item \textit{Idle} for scheduling very low-priority jobs, usually the idle process.
\end{enumerate} 

These classes are hierarchically organized by priority: 
\begin{equation*}
	DL~>~RT~>~CFS~>~IDLE
\end{equation*}
The core scheduler selects which task to run next from the run queue on each core by searching through the scheduling classes in decreasing order of priority.
As a result, the tasks of lower priority classes run in the idle time of higher priority classes. 
We add a new highest priority TT scheduling class ($TT~>~DL$).

%

\section{Implementation}\label{sec:imp}
Let us assume a multicore processor with $N$ CPU cores ($C_0,C_1,\ldots,C_{N-1}$). We let $M$ cores ($C_{N-M},C_{N-M+1},\ldots,C_{N-1}$) out of these $N$ cores ($M<N$) to execute process belonging to the TT class (SCHED\_TT policy). We will refer to them as TT cores. There will be at least one core ($C_0$) that will not run TT processes. We will refer to it/them as non-TT core(s). We also assume that each TT task only has a single thread.

To obtain real-time behavior from the Linux kernel (low latency and high determinism), we chose the fully preemptible kernel (RT) model via kernel configuration during build time. The fully preemptible kernel is available via the PREEMPT\_RT patch. This model ensures that preemption is possible on almost all kernel code apart from some critical sections and implements mechanisms to reduce preemption.
The Linux kernel performs significant asynchronous housekeeping work, such as timekeeping, timer callbacks, interrupt handlers, RCU callbacks, and kernel threads. Using standard Linux kernel configuration, every core is assigned housekeeping work. The ``noise'' from this housekeeping work can significantly impact applications running on the TT cores. 

\begin{figure}[b]
	\centering
	\resizebox{\linewidth}{!} {
		\begin{tikzpicture}[
	ssmirqexec/.style={pattern=grid},
	irqexec/.style={pattern=crosshatch},
	taskexec/.style={pattern=dots},
	slotdiv/.style={line width=0.5mm}
	]
	\def\execheight{0.4}
	\def\ssmirqwidth{0.3}
	\def\irqwidth{0.2}
	\def\slotdivheight{0.6}
	\def\slotlength{2}
	\def\slotstart{0.2}
	\def\slotcount{3}
	\pgfmathsetmacro{\slotlast}{\slotstart + \slotcount*\slotlength}
	\pgfmathsetmacro{\slotsecond}{\slotstart + \slotlength}
	\def\cpux{{0, -1.6, -3.0,-0.6}}
	
	\draw [<-](\slotlast + \slotlength + 0.2,\cpux[0]) -- (0,\cpux[0]) node [left] {$C_0$};
	\draw [<-](\slotlast + \slotlength + 0.2,\cpux[3]) -- (0,\cpux[3]) node [left] {$C_1$};
	\node [left] at (-0.25, {\cpux[1]/2 + \cpux[3]/2}) {\large \rotatebox{90}{. . .}};
	\draw [<-](\slotlast + \slotlength + 0.2,\cpux[1]) -- (0,\cpux[1]) node [left] {$C_{N-M}$};
	\node [left] at (-0.25, {\cpux[1]/2 + \cpux[2]/2}) {\large \rotatebox{90}{. . .}};
	\draw [<-](\slotlast + \slotlength + 0.2,\cpux[2]) -- (0,\cpux[2]) node [left] {$C_{N-1}$};
	
	\foreach \i in {\slotstart,\slotsecond,...,\slotlast} {
		\draw [slotdiv] (\i, \cpux[0]) -- (\i, \cpux[0] + \slotdivheight);
		\draw [slotdiv] (\i, \cpux[1]) -- (\i, \cpux[1] + \slotdivheight);
		\draw [slotdiv] (\i, \cpux[2]) -- (\i, \cpux[2] + \slotdivheight);
		
		\draw [ssmirqexec] (\i,\cpux[0]) rectangle (\i - \ssmirqwidth, \cpux[0] + \execheight);
		\draw [irqexec] (\i,\cpux[1]) rectangle (\i + \irqwidth, \cpux[1] + \execheight);
		\draw [irqexec] (\i,\cpux[2]) rectangle (\i + \irqwidth, \cpux[2] + \execheight);
		\draw [taskexec] (\i+\irqwidth, \cpux[1]) rectangle (\i+\slotlength, \cpux[1] + \execheight);
		\draw [taskexec] (\i+\irqwidth, \cpux[2]) rectangle (\i+\slotlength, \cpux[2] + \execheight);
	}
	
	\pgfmathsetmacro{\slotlengthy}{\cpux[1]/2 + \cpux[2]/2 + \slotdivheight/2}
	\draw (\slotstart, \slotlengthy + 0.2) -- (\slotstart, \slotlengthy-0.2);
	\draw (\slotsecond, \slotlengthy + 0.2) -- (\slotsecond, \slotlengthy-0.2);
	\draw [<->] (\slotstart, \slotlengthy) -- (\slotsecond, \slotlengthy) node [midway, below] {\scriptsize Slot length};
	
	\matrix [draw, below left] at (\slotlast + \slotlength + 0.2, 2.8) {
		\node [draw, ssmirqexec, label=right:Slot shifting manager kernel module HRT interrupt ($C_0$ only)] {}; \\
		\node [draw, irqexec, label=right:SCHED\_TT class HRT interrupt per TT core] {}; \\
		\node [draw, taskexec, label=right: TT processes / VMs / containers] {}; \\
	};
	
\end{tikzpicture}
	}
\caption{SCHED\_TT dispatcher and SSM kernel module}
	\label{fig:TTcores}
\end{figure}
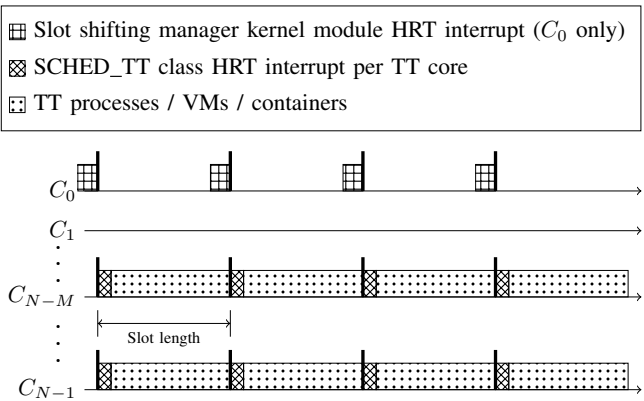

The \textit{isolcpus} parameter helps us to isolate the TT cores from the general SMP balancing and scheduler algorithms.
Similarly, we initialize the $nohz\_full$ to configure $full~dynticks$ along with CPU Isolation, $rcu\_nocb\_poll$ to offload RCU processing to non-TT cores, and the $irqaffinity$ parameter to affine the IRQs to non-TT cores (e.g., $nohz\_full=C_{N-M}, C_{N-M+1},\ldots, C_{N-1}$).
Once the system boots, Linux ensures that no processes execute on these TT cores unless instructed by the slot-sifting manager (SSM) kernel module and restricts all housekeeping work to non-TT cores ($C_0,\ldots, C_{N-M-1}$). Thus, we ensure almost housekeeping noise-free TT cores to run processes assigned to SCHED\_TT policy as depicted in Figure \ref{fig:TTcores}.
After the system boots, we insert SSM and ensure it runs on a non-TT core (e.g., $C_0$). 

The SSM runs as a timer interrupt routine just before the start of each slot. This module runs a partitioned slot-shifting algorithm. It decides what should run in the upcoming slot on each TT core and informs the decision to TT scheduling class via a shared structure.

The TT scheduling class acts as the interface between the SSM and the core Linux scheduler. 
It runs as a periodic timer interrupt routine at the start of every slot. It checks the shared structure set by SSM to check which task to run in the current slot. If the task to run is not the same as the task in the previous slot, it marks the core for rescheduling by the Linux core scheduler. 
The following sections explain the detailed implementation of the TT scheduling class and SSM.

\subsection{TT scheduling class}
We have developed this scheduling class from scratch to implement the SCHED\_TT policy and enable TT scheduling in Linux. We have implemented it on top of Linux kernel version v5.19.9 with PREEMPT\_RT patch.

We defined a macro \textit{SCHED\_TT} in ``include/linux/sched.h'' to declare the new highest priority scheduling class (see Listing \ref{c:macro}). We also add a new variable to the task structure (process descriptor) to act as a logical identifier for a task belonging to the TT class.

\begin{lstlisting}[language=C,basicstyle=\small,frame=single,caption={TT class macro},captionpos=b, label={c:macro},float]
#define SCHED\_DEADLINE 6  // Existing
#define SCHED\_TT       7  // New
struct task_struct {
  ... // Existing task structure
  #ifdef CONFIG_SCHED_TT_POLICY
    unsigned int	TT_id;  // New
  #endif
}
\end{lstlisting}

The information about all ready processes is present in a per-core run queue (RQ) data structure (\textit{struct rq}), which in turn has a TT RQ data structure (\textit{struct tt\_rq tt}) to store information about ready TT tasks on the core (see Listing \ref{c:rqtt}). This structure also has a variable the slot-shifting kernel module uses to indicate to the TT scheduler if it should start running.

\begin{lstlisting}[language=C,basicstyle=\small,frame=single,caption={TT class run queue},captionpos=b, label={c:rqtt},float]
struct rq { // Existing RQ structure
  ...
  struct dl_rq	dl;  // Existing
  #ifdef CONFIG_SCHED_TT_POLICY
    struct tt_rq     tt;	// New
  #endif
}
struct tt_rq {	// New
  ...
  bool scheduler_running;
};
\end{lstlisting}
We made minor additions to the Linux core scheduler code to ensure the kernel knows the new highest priority class and does the additional logging for the TT class if requested. For example, we defined a macro \textit{SCHED\_TT} in ``include/linux/sched.h'' to declare the new highest priority scheduling class. We also add a new variable to the task structure (process descriptor) to act as a logical identifier for a task belonging to the TT class. The information about all ready processes is present in a per-core run queue (RQ) data structure (\textit{struct rq}), which in turn requires a TT RQ data structure (\textit{struct tt\_rq tt}) to store information about ready TT tasks on the core.

The modular Linux scheduler requires each scheduling class to implement some functions specified in the Class structure. Some crucial functions of the TT scheduling class to interface with the core Linux scheduler are show in Listing \ref{c:def}.
\begin{lstlisting}[language=C,basicstyle=\small,frame=single,caption={TT  class definition},captionpos=b,label={c:def},float]
DEFINE_SCHED_CLASS(tt) = {
  // enqueue new runnable TT task in Linux RQ 
  .enqueue_task   = enqueue_task_tt,
  // dequeue stopped TT task from Linux RQ
  .dequeue_task    = dequeue_task_tt,
  // pick next appropriate task to run from
  // the TT scheduling class
  .pick_next_task = pick_next_task_tt,
  ...
\end{lstlisting}

During kernel boot, the core Linux scheduler expects an interface (Listing \ref{c:init}) to initialize the new scheduling class. The initialization of the TT class includes setting up a per-core high-resolution timer (HRT) to have precise timer interrupts at slot boundaries.

\begin{lstlisting}[language=C,basicstyle=\small,frame=single,caption={TT class initialization},captionpos=b, label={c:init},float]
void __init init_sched_tt_class(void){
  // set the tick equal to the slot length
  tt_set_tick_period_ms(TT_TICK_PERIOD);
  ...
  //Set a timer on each TT core
  for_each_possible_cpu(i) {
    tick_timer = per_cpu_ptr(&tt_tick_timer,i);
    hrtimer_init(tick_timer,CLOCK_MONOTONIC,
    HRTIMER_MODE_ABS_HARD);
    //function to call on each tick
    tick_timer->function = tt_cpu_tick;
  }
  ...
}
\end{lstlisting}

The main job of the HRT timer callback (Listing \ref{c:tick}) on each core is to check if the task id assigned by SSM to the current slot for this core differs from the previous one. If it is different, it marks the current core for rescheduling using resched\_curr(). As a result, the Linux core scheduler runs and uses the pick\_ next\_task\_tt() interface to get the pointer to the new task to run. It performs a context switch and then runs the new task in the current slot on that core. In case this interface of the TT class does not provide a task to run, the Linux core scheduler automatically looks for runnable tasks of lower-priority scheduling classes. The same is valid if the TT class tasks finish execution early.
\begin{lstlisting}[language=C,basicstyle=\small,frame=single,caption={TT class CPU tick (per core)},captionpos=b,label={c:tick},float]
static enum hrtimer_restart 
tt_cpu_tick(struct hrtimer *timer){
  int cpu = smp_processor_id();
  ...
  //get the task id of current and next task
  new_sched_tid=this_cpu_read(tt_sched.next);
  prev_sched_tid=this_cpu_read(tt_sched.curr);
  if(new_sched_tid!=prev_sched_tid){
    // only if the previous slot task is not 
    // same as the task for this slot 
    if (new_sched_tid) 
    sched_task = ss_tasks[new_sched_tid];
    else if (sched_task 
             && task_cpu(sched_task)!= cpu){
      //If the task is on different core's RQ,
    } //move it to this one's RQ
    ...
    //mark for resched by Linux core sched
    resched_curr(rq);
  }
  //do not do anything if the task in previous 
  //slot and the current one are the same
  ...
  // restart HRT for the next slot tick
  return HRTIMER_RESTART;
}
\end{lstlisting}

Only non-DL scheduling classes should be used together with the TT class. It is possible to execute new RT aperiodic tasks in the slack of TT tasks via the SSM module.

\subsection{Slot Shifting Manager (SSM) kernel module}\label{sec:impssm}

We assume an external tool handles some things offline, such as precedence constraint resolution of tasks, calculations of the earliest start times and deadlines, and allocation of tasks to cores. The user must provide the information received from the external tool to the SSM kernel module via \textit{sysfs} interface. sysfs is an in-memory file system that allows users to interface with kernel objects such as devices, modules, and other components. 

The SSM sysfs interface allows users to input task count and properties of tasks and perform parts of offline phase specific to the slot shifting algorithm, checking current status, and initiating TT scheduling on TT cores.
Before initiating TT scheduling, the user must use our task execution tool to execute the task binaries as Linux processes and map the Linux process identifiers (PIDs) to the slot-shifting task IDs. The tool sets the scheduling policy for the tasks to SCHED\_TT.

Upon initiation of TT scheduling, SSM performs steps to prepare for slot-shifting and then sets up an HRT timer on core 0 to have precise interrupts before the start of the new slot. Let us assume $WCET_{ssm}$ is the worst-case observed execution time of the SSM module (based on experiments - Section \ref{sec:exp}). The module sets the  HRT timer to cause an interrupt $WCET_{ssm}$ time units before the slot boundary.

The HRT interrupt callback runs a loop to perform the following activity for managing each TT core. We have summarized the SSM HRT callback in Algorithm \ref{alg:ssm}. Firstly, it performs some slot-shifting housekeeping activities such as incrementing slot numbers and updating intervals ($Update\_SlotShifting\_Intervals()$), removing finished tasks, and moving ready periodic tasks to the internal SSM ready queue ($Update\_Interal\_SlotShifting\_RQ()$).
Then, it checks for any new (user-added) periodic or aperiodic tasks that became active. The module must try to add them to the existing scheduling table ($Check\_and\_Add\_New\_RT\_Tasks()$). It performs an acceptance test to check if enough spare capacity is present to add these tasks. If the test is successful, the guarantee routine adds the task temporarily/permanently to the scheduling table and updates the spare capacities of all affected intervals.
Next, it selects the task to run in the next slot of the particular CPU core and updates the internal data structure accessed via the TT scheduling class at the start of each slot ($Set\_Upcoming\_Slot\_Task()$) to find out which task to run in that slot.
Finally, it updates the spare capacity as required ($Update\_SlotShifting\_Spare\_Capacities()$). Since we are running a partitioned slot-shifting algorithm, the HRT callback repeats these steps for each TT core.
\begin{algorithm}[t]
	\caption{SSM HRT timer callback}
	\label{alg:ssm}
	\begin{algorithmic}
		\Function{ssmod\_timer\_cb(struct hrtimer *time)}{}
		\State $core := N-M$
		\While{$core < N$}
		\State $Update\_SlotShifting\_Intervals()$
		\State $Update\_Interal\_SlotShifting\_RQ()$
		\State $Check\_and\_Add\_New\_RT\_Tasks()$
		\State $Set\_Upcoming\_Slot\_Task()$
		\State $Update\_SlotShifting\_Spare\_Capacities()$
		\State $core := core + 1$
		\EndWhile
		\EndFunction
	\end{algorithmic}
\end{algorithm}
\section{Experiments}\label{sec:exp}
We evaluated our implementation on a Dell COTS server with an Intel Xeon processor (Cascade Lake, 16 physical CPU cores, $2.3GHz$). We turned off certain configurations, such as power-saving mechanisms (e.g., frequency scaling, C-states) and hardware multi-threading, in BIOS and Linux Kernel to avoid unpredictability sources for the TT cores.
Using power-saving mechanisms such as low power stats leads to higher scheduling overhead, while frequency scaling leads to an increase in the WCET of tasks. Kadusale et al.~\cite{kadusale} present an energy-aware slot-shifting algorithm variation that considers lower power states and frequency scaling. However, we do not consider energy-aware slot-shifting in this paper.
We ran the slot-shifting manager (SSM) on core 0 and
isolated cores 1 to 15 to run offline scheduled TT tasks as explained in Section \ref{sec:imp}.

We specify parameters (Table \ref{tab:ohead_exp_tset_params}), such as the total target utilization, WCET range, and period range, to the UUnifast algorithm \cite{bini2005measuring} to generate task sets for evaluation.
Each task in every task set performs many arithmetic operations (in a loop) and uses the clock cycle performance counter event to determine the time needed to perform all the operations in each iteration of the loop. Offline scheduled tasks have total utilization of ca. 50\%. New AP RT tasks that the slot-shifting manager must add during runtime have an additional total utilization of 50\%. 
\begin{table}[ht]
	\centering
	\begin{tabular}{ c|c|c} 
		Parameter & Offline tasks & AP RT tasks\\
		\hline
		WCET range & [1,15] & [10,15] \\ 
		Period range & [15,50] & [10,15]\\
		Total Taskset Utilization & 50\% & 50\% \\
	\end{tabular}
	\caption{Taskset Parameters}
	\label{tab:ohead_exp_tset_params}
\end{table}

We created 50 task sets via the UUnifast algorithm for the evaluation. We parse and feed the UUnifast algorithm output to the slot-shifting manager via the sysfs interface (see Section \ref{sec:impssm}). Each task set has an offline schedule duration between 480 to 520 slots, with a slot length of 3ms. We ran each test 5 times. Thus, we measure the results in the following sections for scheduling overheads by the TT class and the slot-shifting manager by considering more than $120\times 10^3$ slots $\times 15$ cores = $18 \times 10^5$ slots in total. The main aim of these task sets is not to test the working for the slot-shifting algorithm (see previous work \cite{schorr2010online}) but to measure the worst-case overheads for SCHED\_TT and estimate the overheads for SSM. We do not expect a change in SCHED\_TT overheads with different utilization values.

\subsection{TT class overheads}

\begin{figure}[h]
	\centering
	\includegraphics[width=0.9\columnwidth]{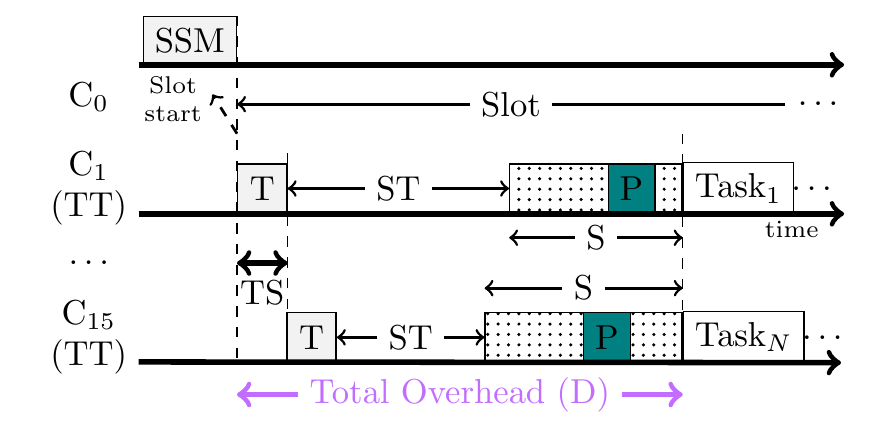}
	\caption{TT class overhead components}
	\label{fig:overhead}
\end{figure}

This section shows observed periodic overheads of the individual components of the TT scheduling class and its interaction with the Linux core scheduler. We measured all the overheads by reading elapsed clock cycles (PMU events) at the start and end of the appropriate code sections. These components are depicted in Figure~\ref{fig:overhead}.
Task Tick (T) represents the overhead for the TT class HRT callback (Listing \ref{c:tick}).
Tick Skew (TS) represents the maximum difference in occurrence time of HRT time callback on each core.
Schedule Trigger (ST) is the time needed for the Linux core scheduler to activate once the TT class HRT callback marks a core for rescheduling. 
$\_\_schedule$ function (S) is the primary function of the Linux core scheduler. This function is triggered on TT cores only when the TT class marks a core for rescheduling. It puts the previously running task into the appropriate RQ, calls TT class' $pick\_next\_task$, and lastly, performs a context switch before passing the control to the newly selected task.
$pick\_next\_task$ function (P) is called by $\_\_schedule$ to pick a new task of the highest priority scheduling class to run next. In our case, the highest priority scheduling class is SCHED\_TT. $pick\_next\_task$ in turn calls $pick\_next\_task\_tt$ (Listing \ref{c:def}).

\begin{figure}[tb]
	\centering
	\begin{subfigure}{0.8\linewidth}
		\caption{TT task\_tick (T)}
		\includegraphics[width=\columnwidth]{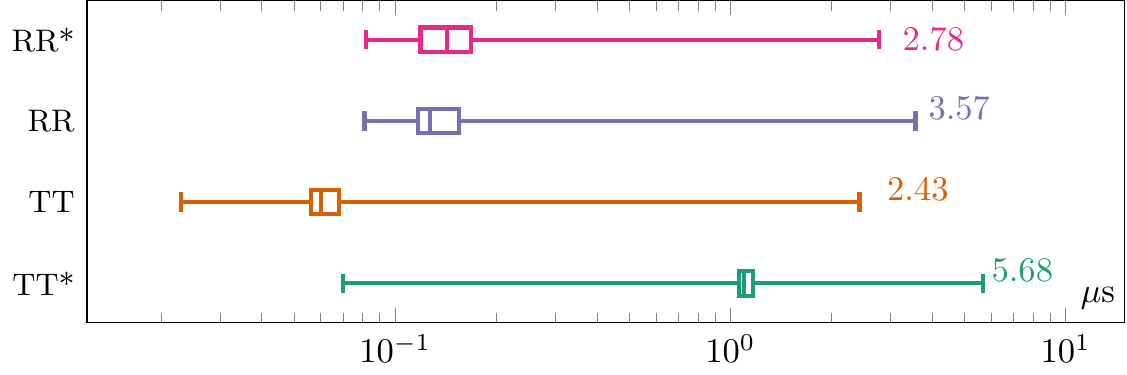}
		\label{sfig:ohead_tick}\vspace{-0.5cm}
	\end{subfigure}
	\begin{subfigure}{0.8\linewidth}
		\caption{Schedule trigger (ST)}
		\includegraphics[width=\columnwidth]{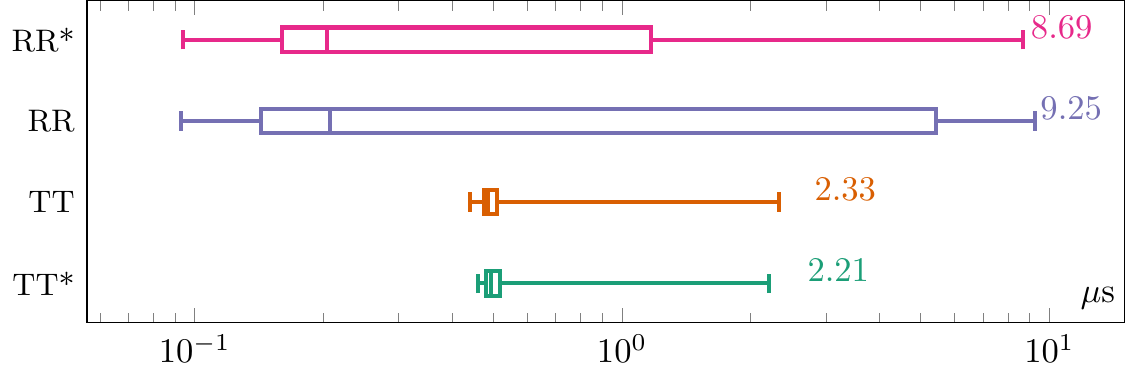}
		\label{sfig:ohead_st}\vspace{-0.5cm}
	\end{subfigure}
	\begin{subfigure}{0.8\linewidth}
		\caption{Linux \_\_schedule (S)}
		\includegraphics[width=\columnwidth]{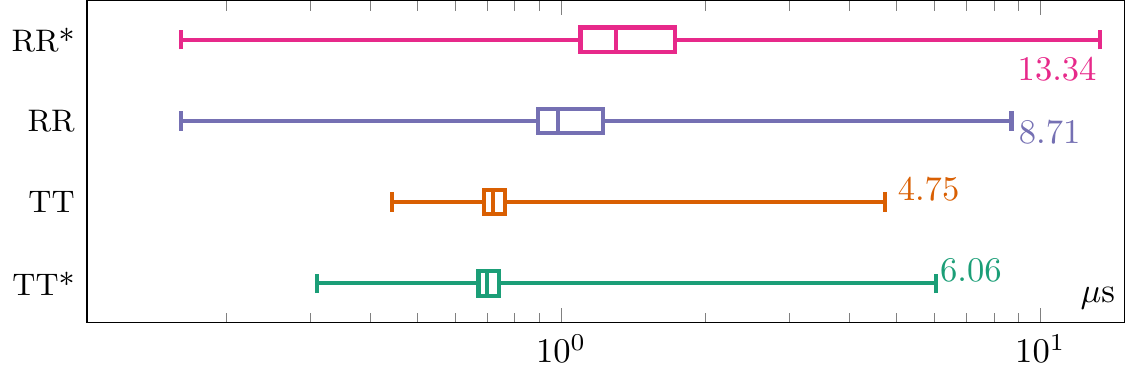}
		\label{sfig:ohead_sched}\vspace{-0.5cm}
	\end{subfigure}
	\begin{subfigure}{0.8\linewidth}
		\caption{TT pick\_next\_task (P)}
		\includegraphics[width=\columnwidth]{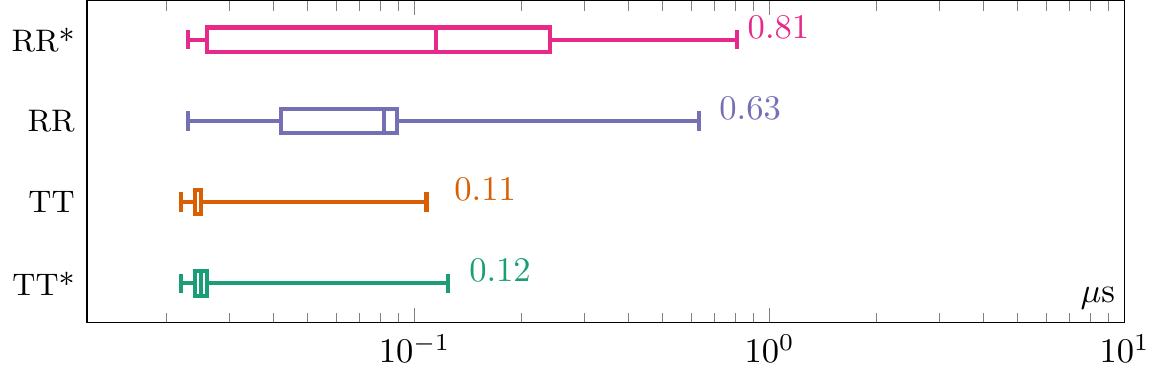}
		\label{sfig:ohead_pick}\vspace{-0.5cm}
	\end{subfigure}
	\begin{subfigure}{0.8\linewidth}
		\caption{Tick Skew (TS)}
		\includegraphics[width=\columnwidth]{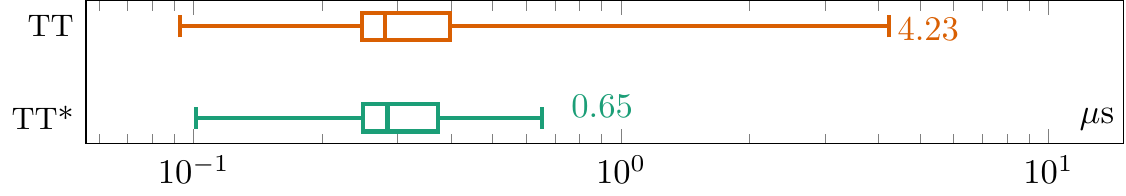}
		\label{sfig:ohead_tickskew}\vspace{-0.5cm}
	\end{subfigure}
	\begin{subfigure}{0.8\linewidth}
		\caption{Total duration (D)}
		\includegraphics[width=\columnwidth]{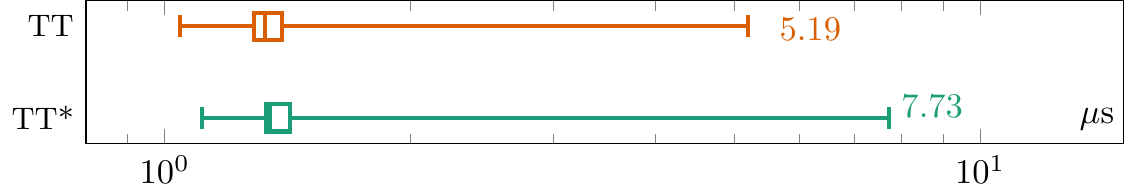}
		\label{sfig:ohead_tot}\vspace{-0.5cm}
	\end{subfigure}\vspace{-0.25cm}
	\caption{Overhead measurements for TT class  and RT class (RR) without and with(*) core migration}
	\label{fig:ohead_mcores}
\end{figure}

We perform experiments with offline schedules that require and do not require migration of tasks between cores. We also compare our results to similar components from the existing RT scheduling class (SCHED\_RR round robin policy), which is closest to TT class style scheduling. Results are shown in Figure~\ref{fig:ohead_mcores}. TT class performs slightly better for task tick without migration. However, the TT class performs slightly worse with migrations because we artificially generated an offline schedule that causes multiple migrations in the same slot. A similar schedule is challenging to simulate for SCHED\_RR as it does task migration only for load-balancing purposes. 
We observed a maximum tick skew of $4.23\mu s$ for the TT class tick. We cannot make a meaningful comparison since RT class ticks are not precise.
We observed lower overheads for the schedule trigger and $\_\_schedule$ function for the TT class. We expected little difference as most code (except some interfaces) is part of the Linux core scheduler. We observed a significant difference in $pick\_next\_task$, where the TT class has much lower overheads.

The maximum observed total overhead (D) for the TT class with and without migrations was $7.73\mu s$ and $5.19\mu s$ in the worst case, with more than 99\% slots requiring $<2.66 \mu s$ and $<2.47 \mu s$, respectively.
We targeted a slot duration of $3ms$. We read the clock cycles at the start of each slot and observed a maximum slot duration of $3.003ms$, a minimum slot duration of $2.995ms$, and an exact $3ms$ duration for $>99\%$ slots.

\paragraph*{\textbf{Comparison to existing work}}
Karachatzis et al.~\cite{schedTTech} experimentally demonstrate that their kernel-level TT scheduling implementation has lower scheduling latency ($max = 41.79 \mu s$) as compared to the existing completely fair scheduling policies in Linux ($max = 12813.56 \mu s$) and MotionWise~\cite{motionwise} userspace TT implementation ($max = 85.67 \mu s$). The experiments were conducted on a quad-core Intel Atom processor ($1.594 GHz$) on Linux kernel v5.9.1 with PREEMPT\_RT and optimizations for running RT tasks. We did not observe further details about how the scheduling latency was precisely determined to make a clear comparison. In our implementation, we observed a worst-case scheduling latency of $5.19 \mu s$ (over $18 \times 10^5$ slots) with migration disabled (similar to them) on server-grade hardware (16 cores, $2.3 Ghz$), Linux kernel v5.19.9, and similar patch and optimizations. We measured the latency by reading hardware clock cycles at appropriate time points within the Linux core scheduler. 

\subsection{SSM kernel module overhead}
\begin{figure}[t]
	\centering
	\includegraphics[width=0.8\columnwidth]{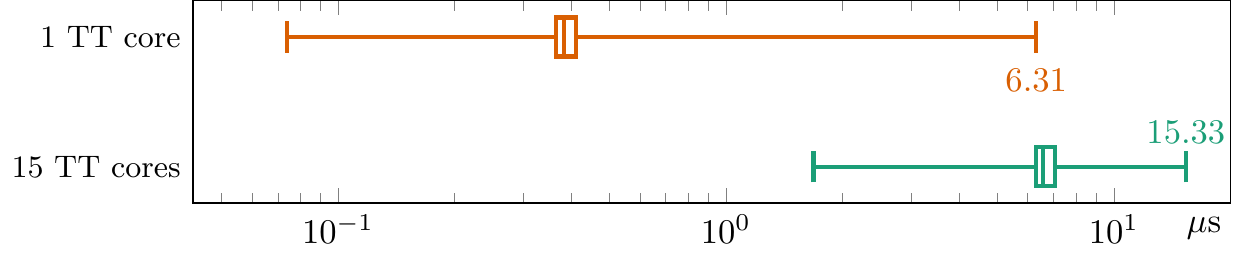}\vspace{-0.25cm}
	\caption{SSM kernel module overhead}
	\label{sfig:ohead_ssm}
\end{figure}

\begin{figure}[t]
	\centering
	\begin{subfigure}{0.9\linewidth}\centering
		\includegraphics[width=\columnwidth]{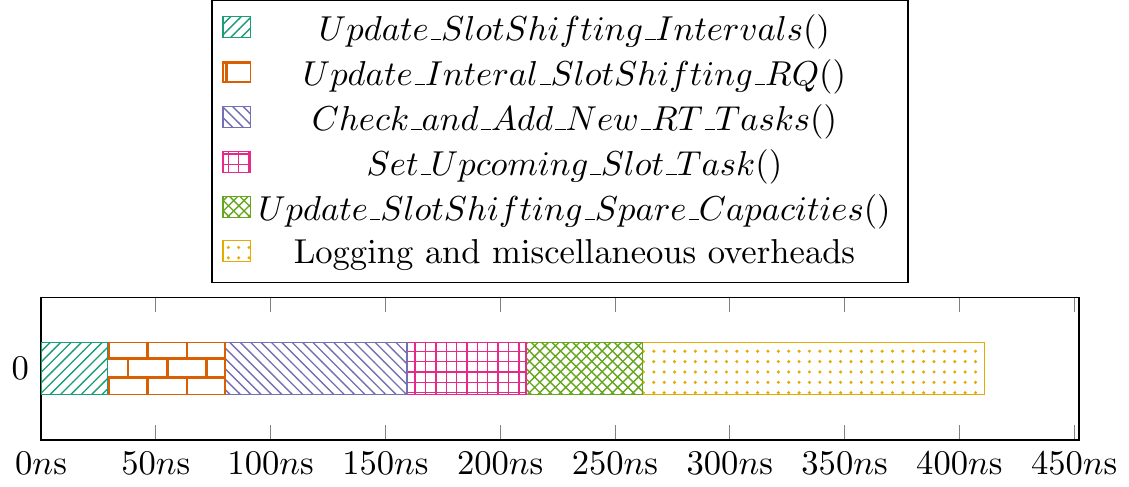}\vspace{-0.15cm}
		\caption{Average}
		\label{fig:ohead_ssm_break_avg}\vspace{0.15cm}
	\end{subfigure}
	\begin{subfigure}{0.9\linewidth}\centering
		\includegraphics[width=\columnwidth]{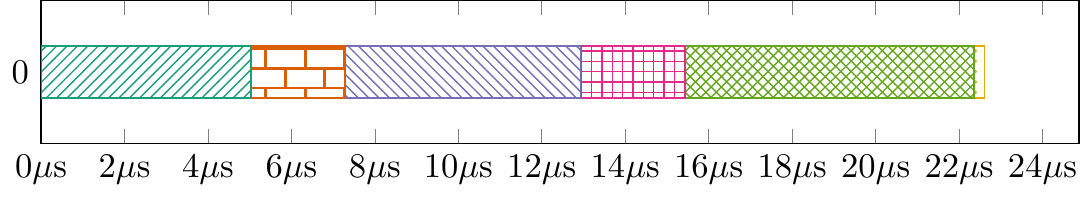}\vspace{-0.15cm}
		\caption{Worst case}
		\label{fig:ohead_ssm_break_wc}
	\end{subfigure}\vspace{-0.15cm}
	\caption{Overheads of individual SSM functions}
	\label{fig:ohead_ssm_break}
\end{figure}

We measured the overheads for the SSM module by reading elapsed clock cycles (PMU events) at the start and end of the module.
Figure~\ref{sfig:ohead_ssm} shows the observed periodic overheads of the SSM module running on core 0 and executing the slot-shifting algorithm for one single core and all 15 cores together. 
We also measured the overheads for the individual functions from Algorithm \ref{alg:ssm} by reading elapsed clock cycles (PMU event) at the start and end of the functions.
Figure~\ref{fig:ohead_ssm_break} shows the worst and average overheads of the individual functions.
Similar overheads for slot-shifting algorithms were measured and explained by Schorr et al.~\cite{schorr2010online}. Hence, we do not explain these results in detail in this paper.


Note that the worst case seems longer in Figure~\ref{fig:ohead_ssm_break_wc} as compared to Figure~\ref{sfig:ohead_ssm} since Figure~\ref{fig:ohead_ssm_break_wc} shows the cumulative worst case of individual functions.
We use the worst-case observed value ($15.33 \mu s$) to set up the HRT timer interrupt for SSM, as explained in Section \ref{sec:impssm}. 

We can observe that only a tiny portion of the overhead goes into interfacing with the TT class (to set tasks for upcoming slots). This indicates a possibility to integrate and test other joint TT and ET scheduling algorithms in the future instead of just allowing slot-shifting algorithms.
%
%
%
\section{Conclusion}
We presented a new scheduling class for the Linux scheduler to support time-triggered (TT) scheduling of tasks on multicore Linux nodes. We complemented the new scheduling class with a low overhead slot-shifting manager (SSM) running on a non-TT core to provide guaranteed execution time to real-time aperiodic tasks by using the slack of the time-triggered tasks and avoiding a high-overhead table regeneration for adding new periodic RT tasks. We have implemented the TT class for Linux kernel v5.19.9 with PREEMPT\_RT patch and evaluated our implementation on server-grade hardware with Intel Xeon Scalable Processor. Our implementation has very low overheads and performs better than the existing RT class (SCHED\_RR), especially for picking tasks to run next. We also observed that more than 99\% of slots achieve the targeted slot length with a maximum error of $5ns$ in the rest.
We observed extremely low worst-case TT class scheduling ($<7.73\mu s$) and SSM overheads ($<15.33\mu s$).

We are fine-tuning the implementation at present and working on making it open-source. In the future, we want to support multi-threaded / multicore tasks (process, VMs, and containers).
In this paper, we focused on TT CPU scheduling alone. However, other sources of unpredictability exist, such as the memory hierarchy and network.
Thus, we want to consider memory bandwidth and cache allocation to slots and explore integration with time-sensitive networking (TSN).
We aim to make the SSM modular more generalized to allow integration of other joint time-and event-triggered algorithms. 
We leave the integration of TT Linux nodes in a distributed system or cloud to future work. It will require some form of clock synchronization with other nodes.
Since embedded devices have a limited number of cores, we want to explore the possibility of avoiding non-TT cores by integrating SSM with SCHED\_TT class.

\section*{Acknowledgment}
We want to thank all the anonymous reviewers for their time and expertise in providing feedback on our paper. We highly appreciate their invaluable inputs and support.

\bibliographystyle{IEEEtran}      
\bibliography{references}   

\end{document}